\documentclass[pra,twocolumn,psfig]{revtex4}
\usepackage{bm}
\usepackage{epsfig}

\begin{document}
\def\bea{\begin{eqnarray}} 
\def\eea{\end{eqnarray}} 
\def\be{\begin{equation}}
\def\ee{\end{equation}} 
\def\rra{\right\rangle} 
\def\lla{\left\langle}
\def\ra{\rightarrow} 
\def\bp{\bm{p}}
\def\bP{\bm{P}}

\title{Deformed Fermi surfaces in ultracold Fermi gases}

\author{Armen Sedrakian$^1$\footnote{armen.sedrakian@uni-tuebingen.de}
}
\author{Jordi Mur-Petit$^2$\footnote{jordim@ecm.ub.es}
}
\author{Artur Polls$^2$\footnote{artur@ecm.ub.es}
}
\author{Herbert Müther$^1$\footnote{herbert.muether@uni-tuebingen.de}
}

\affiliation{ 
$^1$Institut für Theoretische Physik, 
Universität Tübingen, D-72076 Tübingen, Germany \\
$^2$Departament d'Estructura i Constituents de la Matèria,
Universitat de Barcelona, E-08028 Barcelona,
    Spain
   }

\date{\today}

\begin{abstract}
The superfluid transition in an ultracold two-component atomic Fermi gas is analyzed
in the case where the two components have different densities. We describe a superfluid 
state which spontaneously breaks the rotational-symmetry by deforming the Fermi surfaces of
both species into ellipsoidal form. At relatively large hyperfine-spin asymmetries, this 
deformation is shown to help the appearance of pairing, which in the rotationally-symmetric 
(BCS) case would be forbidden by Pauli blocking. The prospects for experimental detection of 
such a deformed Fermi surface phase are discussed.
\end{abstract}

%\bigskip
\maketitle

\tighten

The present capabilities of cooling of atomic ensembles allow for reasonable
expectations to observe a superfluid transition in ultracold fermionic
systems, in direct analogy to the BCS superconductivity
\cite{STOOF}. The temperatures achieved in recent experiments on 
fermionic atoms~\cite{MOLECULAR_BEC} are a fraction ($\sim 0.1-0.3$) of the
Fermi-temperature, i.e.  atoms in a trap are in the quantum degenerate
regime and, therefore, 
attractive two-body forces are expected to drive the Cooper instability.
The strength of the two-body interactions can be tuned using a Feshbach 
resonance by varying the external magnetic field~\cite{FESHBACH},
thus the entire range from weak to strong
couplings can be probed. In the crossover region the Feshbach
resonance may strongly enhance the pairing interaction and give rise to high
temperature superfluidity~\cite{RESONANT_PAIRING}.
Recent experiments have probed the condensation of fermionic pairs above the 
Feshbach resonance, where the system does not support a genuine two-body
bound state~\cite{JILA04,MIT04,NC04}. Whether the observed correlated 
pairs are, in fact,  weakly bound and spatially extended Cooper pairs, 
is not clear yet; however the measured collective modes of $^6$Li
atoms under these conditions are consistent with the superfluid
hydrodynamics of a Fermi-gas and provide evidence for superfluidity 
in a resonantly interacting Fermi gas \cite{NC04}.

The very low temperatures (in the nanokelvin range) and densities
reached in the experiments  considerably reduce the contribution from 
$L\neq0$ orbital angular momentum waves to atomic collisions. 
Therefore, $s$-wave collisions, which can be characterized by 
the scattering length $a$,  are the most relevant for the description 
of these systems. As usual, we take $a<0$ to indicate an attractive 
interaction between the atoms.
Since Pauli's exclusion principle forbids $s$-wave 
interaction between indistinguishable fermions the pairing should appear
between fermionic atoms belonging to different hyperfine states. 
Such systems, where two hyperfine levels are populated,
have been created and studied experimentally with $^6$Li and $^{40}$K
atoms~\cite{MOLECULAR_BEC,JILA04,MIT04,NC04}. 
The BCS theory predicts a suppression of the pairing
correlations when  the Fermi-energies, or equivalently, the densities 
of the two hyperfine states $\vert 1\rangle$ and $\vert 2\rangle$ are
different. In the low density limit $k_F|a|
\ll 1$ ($k_F$ is the Fermi-momentum) the value of the critical
density  asymmetry $\alpha = (\rho_1-\rho_2)/(\rho_1+\rho_2)$, 
where $\rho_{1,2}$ are the
densities of hyperfine states $\vert 1\rangle$ and $\vert 2\rangle$, 
for which the superfluidity vanishes can be deduced analytically~\cite{PLA}.
The dependence of the pairing gap $\Delta$ at the Fermi surface on the 
total density $\rho=\rho_1+\rho_2=k_F^3/(3\pi^2)$ 
and the density asymmetry $\alpha$ is described by 
\begin{equation}
\label{eq:BCSASYM}
  {\Delta (\alpha) \over \Delta_0} 
   = \sqrt{1-{4\mu \over 3\Delta_0}\alpha} ,
\end{equation}
where 
$\Delta_0\simeq {8 e^{-2}}\mu\exp{\left[-{\pi / (2k_F|a|})\right]} \ll 1$ 
is the gap in the symmetric matter and $\mu$ is the chemical potential.
Therefore the gap disappears for asymmetries 
$\alpha >\alpha_{\rm max}=3\Delta_0/(4\mu)$,
which in this limit is a very small number. For example, 
for the pairing of $^6$Li atoms in the states 
$|1\rangle=|F=3/2,m_F=3/2\rangle$ and
$|2\rangle=|3/2,1/2\rangle$, for which the triplet scattering length 
is $a=-2160a_B$ (where $a_B$ is the Bohr radius),
and at the density $\rho=3.8\times10^{12}$ cm$^{-3}$ (corresponding to
$k_F|a|=0.55$) the maximum asymmetry at which BCS pairing is possible is only 
$\alpha_{\rm   max}=0.07$.

The purpose of this work is to demonstrate that the superfluid state
in ultracold atomic gases can persist 
for density asymmetries $\alpha > \alpha_{\rm  max}$, 
and can be enhanced in a range of  
$\alpha < \alpha_{\rm  max}$ due to spontaneous deformation 
 of the Fermi spheres of two-hyperfine states in the momentum space. It has
been shown earlier (in other contexts) that the deformation of the
Fermi surfaces of asymmetric two-component superconductors into ellipsoidal form
leads to a novel ground state with deformed Fermi surfaces; the
associated superconducting phase brakes global rotational 
symmetry of the space from O(3) down to O(2)~\cite{DFS_PRL}. 
The deformed Fermi surface superfluid (DFS) 
phase belongs to the class of the superconducting
states with broken space symmetries, an example of which is the 
Larkin-Ovchinnikov-Fulde-Ferrell (LOFF) phase \cite{LOFF}; which has
been studied in ultracold atomic gases in Ref.~\cite{COMBESCOT}.
The underlying principle that makes these phases favorable is the
compensation in the increase in the kinetic energy of the
superconducting phase 
by the gain in the (negative) condensation energy due to a rearrangement
of quasiparticle distributions. In the 
LOFF phase the compensation is achieved by
sampling Cooper pairs with finite total center-of-mass momentum; in
the DFS phase the same is achieved by a deformation of Fermi surfaces
 at zero total momentum of
the pairs. Formally, these phases correspond to the first and 
second order expansion of the quasiparticle spectrum with respect to
the angle formed by the particle momentum and the axis of spontaneous 
symmetry breaking.   

Consider a uniform gas of Fermi atoms with two hyperfine states, which 
we assign labels 1 and 2 (these states equivalently can be thought of as 
pseudospins $\uparrow$ and $\downarrow$.) The model Hamiltonian that 
describes our system is 
\be\label{HAMILTON}
\hat H = \sum_{\bp,\sigma} \epsilon_{\bp} \hat a^{\dagger}_{\bp\sigma} 
\hat a_{\bp\sigma} 
- g\sum_{\bp\bp'} \hat a^{\dagger}_{\bp',1}  \hat a^{\dagger}_{-\bp',2} 
\hat a_{-\bp,2} \hat a_{\bp,1} ,
\ee
where $\hat a^{\dagger}_{\bp\sigma}$ and $\hat a_{\bp\sigma}$ are the
creation and annihilation operators of a state with momentum $\bp$,
pseudospin $\sigma (= 1,2)$ and energy $\epsilon_{\bp} =p^2/2m$,
where $m$ is the atom bare mass (here and below we set the volume $V =1$). 
The two-body coupling constant is determined by
the $s$-wave scattering length $a<0$ as $g = 4\pi\hbar^2\vert a\vert/m$.
In the following we will work in a scheme where the particle number
conservation is explicitely implemented by fixing their densities
$\hat \rho_{1 (2)} = \sum_{\bp} \hat n_{\bp, \sigma} $, $\hat n_{\bp, \sigma} 
= \hat a^{\dagger}_{\bp,1 (2)}  \hat
a_{\bp,1 (2)}$ (or equivalently the total density $\rho$ and the
asymmetry parameter $\alpha$) by adjusting the chemical potentials 
$\mu_{\sigma}$ of the hyperfine states. There is a significant difference
between the scheme above and the one where the total chemical
potential is fixed and the gap is studied as a function of the
difference in the chemical potentials of species; in the latter case
double valued solutions appear which are absent in 
the former case~\cite{BCS_ASY}.

The mean-field solutions for the model Hamiltonian
(\ref{HAMILTON}) can be obtained by  diagonalizing 
the Hamiltonian with the help of the familiar
Bogolyubov transformations: $\hat b_{\bp, 1} = u_p\hat a_{\bp, 1}
+v_p\hat a^{\dagger}_{-\bp, 2}$ and $\hat b_{\bp, 2} 
= u_p\hat a_{\bp, 2}-v_p\hat a^{\dagger}_{-\bp,  1}$, where 
$u_{\bp}^2+v_{\bp}^2 =1$. 
A variational minimization of the energy with respect to the
parameter $u_{\bp}$ (or $v_{\bp}$) leads to the gap equation 
\be\label{GAP}
\Delta = g\int\!\! 
\frac{d \bp}{(2\pi)^3}\,  u_{\bp} \, v_{\bp}             
\left[1-f(E_1)-f(E_2)\right],
\ee
where  $f(E) = \left[1+{\rm exp}(E/T)\right]^{-1}$ is the Fermi
distribution function, $T$ is the temperature. The two branches 
of quasiparticle spectra are defined as 
\be\label{QP} 
\left. \begin{array}{cc} E_1 \\
E_2 \end{array}\right\}= \sqrt{\xi_S^2+\Delta^2}\pm \xi_A ,
\ee
where the symmetrized 
$\xi_S = \frac{1}{2}\left(\varepsilon_1+\varepsilon_2\right)$
and anti-symmetrized 
$\xi_A = \frac{1}{2}\left(\varepsilon_1-\varepsilon_2\right)$ spectra  
are written in terms of the normal state spectra
$\varepsilon_{\sigma} = \epsilon_{\bp} -\mu_{\sigma}$  
(we do not distinguish the masses of different hyperfine states) and  
the transformation  parameters are defined as   
\begin{eqnarray}
\left. \begin{array}{cc} u_{\bp}^2 \\
v_{\bp}^2 
\end{array}\right\}&=& {1\over 2}
\left(1 \pm {\xi_S\over \sqrt{\xi_S^2+|\Delta |^2}}\right).
\end{eqnarray}
The occupation of the states in the superfluid phase are given by           
\be\label{OCCUP} 
n_{\bp, 1(2)} = \left\{u_{\bp}^2f(E_{1(2)}) 
+ v_{\bp}^2[1-f(E_{2(1)})]\right\},
\ee 
with the normalization condition
\be\label{NORMALISATION} 
\rho_{\sigma} = \sum_{\bp} n_{\bp,  \sigma}.
\ee

We now turn to the description of the perturbations of the Fermi surfaces from the
spherically symmetric form and the study of the stability of these
perturbations.  The two Fermi surfaces in momentum space are defined by
the equations $\varepsilon_{\sigma} = \epsilon_{\bp} -\mu_{\sigma} =
0$. When the chemical potentials 
$\mu_{\sigma} = p^2_{F, \sigma}/2m$, where $p_{F, \sigma}$ are
the Fermi-momenta of the hyperfine states, are isotropic in the
momentum space the Fermi surfaces are spherical. Relaxing the latter
assumption we expand the quasiparticle spectrum in spherical harmonics
$\varepsilon_{\sigma} = \sum_l \varepsilon_{l \sigma} P_l(x)$, where $x$
is the cosine of the angle formed by the particle momentum 
and a randomly chosen symmetry breaking axis, 
$P_l(x)$ are the Legendre polynomials. 
The $l = 1$ terms break the translational symmetry
by shifting the Fermi surfaces without deforming them; these terms are 
ignored below. Truncating the expansion at the second order ($l =
2$), we rewrite the spectrum in a form equivalent to the above one \cite{DFS_PRL}
\begin{equation}
\varepsilon_{\sigma} = \epsilon_{\bp} 
-\mu_{\sigma}\left(1+\eta_{\sigma} x^2\right) ,
\end{equation}
where the parameters $\eta_{\sigma}$ describe the quadrupole deformation of the
Fermi surfaces. It is convenient to work with the symmetrized and
anti-symmetrized combinations $\delta\epsilon  = (\eta_1-\eta_2)/2$ and 
$\Xi = (\eta_1+\eta_2)/2$, which we will refer to
as `relative' and `conformal' deformations. Our next task is
to examine the energy of the superfluid state at finite
deformations to assess whether the deformations lower the energy of
the system and lead to a new stable ground state. We shall work at
fixed temperature and  number density of the hyperfine states and
will examine the difference between the 
free-energies of the superfluid state with
deformations and the undeformed normal state. We shall assume that the
conformal deformation is absent $\Xi = 0$ and look for a minimum of
this difference with respect to  a single parameter
$\delta\epsilon$. (The situation is similar to the description of the 
LOFF phase where one allows for the normal state spectrum to have a finite 
momentum $\bP$ and  seeks the minimum of the appropriate
thermodynamical potential as a function of $\bP$).

The free-energy of the superfluid phase is defined as
\be
{F}_S = E_{\rm kin}+E_{\rm pot} - TS_S,
\ee
where the first two terms comprise the internal energy which is the
statistical average of the Hamiltonian (\ref{HAMILTON}) and  $S_S$ 
is the superfluid entropy, defined by the well-known combinatorial
expression. In our model the sum of the kinetic 
and potential energies is  
\be 
E_{\rm kin}+E_{\rm pot} = \sum_{\bp} \epsilon_{\bp}
\left(n_{\bp,1}+n_{\bp,2}\right) -\frac{\Delta^2}{g}.
\ee
The free-energy of the undeformed normal state follows by setting in
the  above expressions $\Delta = 0 =\delta\epsilon$. 
Because of the contact form of the interaction the gap equation and
the superfluid kinetic energy need a regularization. The regularized
gap equation is
\be\label{GAP2}
1 = \frac{g}{2(2\pi)^2}\int_0^{\Lambda}\!\! dp p^2\!\!
\int_{-1}^{1}\!\! dx 
\left(\frac{1-f(E_1)-f(E_2)}{\sqrt{\xi_S^2+\Delta^2}}- 
\frac{\gamma}{\epsilon_{\bp}}\right),
\ee
where the case $\gamma =1$ and $\Lambda \to \infty$ corresponds the
common practice of regularization \cite{STOOF}, which
combines the gap equation with the $T$-matrix equation in the free
space. The case $\gamma =0$ and finite $\Lambda$ corresponds to 
the cut-off regularization of the original gap equation. The term
$E_{\rm kin}$ is regularized by a cut-off $\Lambda$, which is deduced 
from the requirement that both regularization schemes give the same
result for the gap.

\begin{figure}[t] % fig 1
\epsfig{figure=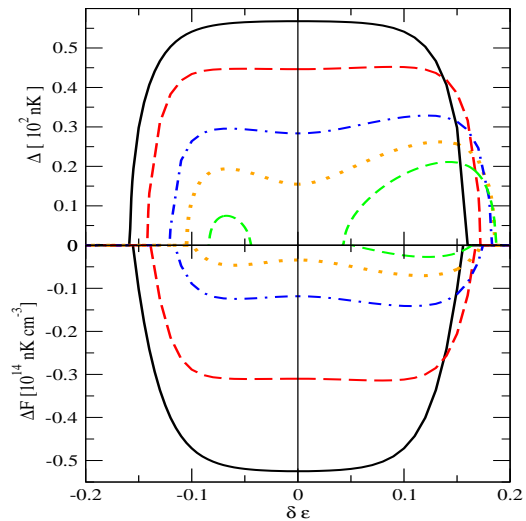,height=16cm,width=9.cm,angle=-90}
\caption{(Color online)
  The dependence of the pairing gap (upper panel) and the free-energy 
  difference (lower panel) on relative deformation for $k_F a = 0.55$,
  at temperature $T = 10$ nK, density  $\rho=3.8\times10^{12}$
  cm$^{-3}$, and constant $\alpha = 0.0$ (solid line), $\alpha = 0.02$ 
  (dashed line), $\alpha = 0.04$ (dashed-dotted line),
  $\alpha = 0.05$ (dotted line) and $\alpha = 0.057$ (short-dashed line).
}
\label{MSfig:fig1}
\end{figure}

Eq. (\ref{GAP2}) was solved numerically with the 
constraint (\ref{NORMALISATION}) for various values of the dimensionless
parameter $k_F a$ at density $\rho=3.8\times10^{12}$ cm$^{-3}$ and 
temperature $T=10$ nK. This density corresponds to  $k_F\approx 
4.8\times10^4$ cm$^{-1}$ and Fermi-temperature $T_F = 942$  nK.
The triplet scattering length in vacuum for $^6$Li atoms 
in the hyperfine states  $|1\rangle=|F=3/2, ~m_F=3/2\rangle$ and 
$|2\rangle=|3/2, ~1/2\rangle$ is $a=-2160a_B$, but as 
already pointed out can be easily manipulated 
using Feshbach resonances.  Figure~\ref{MSfig:fig1} displays the dependence
of the pairing gap and the free-energy difference 
$\Delta F = F_S - F_N$  on the relative
deformation for several density asymmetries and zero conformal deformation.
We restrict the density asymmetry to positive values, i.e. assume 
$\rho_1 > \rho_2$; positive values of $\delta\epsilon$ correspond 
to a prolate (cigar-like) deformation of the majority and oblate
(pancake-like) deformation of the minority population's Fermi-spheres; 
for negative $\delta\epsilon$ the  
reversed is true. When there are no deformations, the antisymmetric 
part of the quasiparticle spectrum (\ref{QP}), $\xi_A$, acts in 
the gap equation (\ref{GAP}) to reduce the phase space coherence between 
the quasiparticles that pair; (when $\xi_A = 0$ the BCS limit is recovered 
with equal occupations for both particles and perfectly matching 
Fermi surfaces). This blocking effect is responsible for the reduction 
of the gap with increasing asymmetry and its disappearance above 
$\alpha \simeq 0.07$ [see Eq. (\ref{eq:BCSASYM})]. 
Allowing for deformations introduces a modulation of  
$\xi_A$ with the cosine of the polar angle 
$x$ (in the frame where the $z$ axis is 
along the symmetry breaking axis), which 
acts to restore the phase space coherence for some values of $x$ 
at the cost of even lesser coherence for the remainder values.
The result, seen in Figure~\ref{MSfig:fig1}, is the {\it increase}
of the gap for finite deformations. At extreme large asymmetries the 
re-entrance effect sets in: the gap exists only for the deformed state, 
with lower and upper critical deformations marking the pairing regions.
These features are seen for both the positive and negative values 
of $\delta\epsilon$, but are more pronounced for $\delta\epsilon>0$;
(quite generally our equations are not symmetric under the interchange
of the sign of deformation, except when $\alpha = 0$).
The free-energy difference $\Delta F$ mimics basically 
the gap function due to the contribution from the potential
energy; note that the critical values of deformations 
at which $\Delta F$ vanishes do not coincide with those for the 
gap due to the positive contribution of the kinetic energy difference.
The same calculations as above were carried out for larger couplings
$k_Fa = 1$ and $k_Fa = 2$ with qualitatively similar results; 
the gaps found in the symmetric case are 1.93 and 3.75 nK, respectively,
the reentrance effect is observed in  each case
for asymmetries around 0.18 and 0.3
and the pairing disappears above the asymmetries 0.22 and 0.43.
\begin{figure}[bht] % fig 2
\epsfig{figure=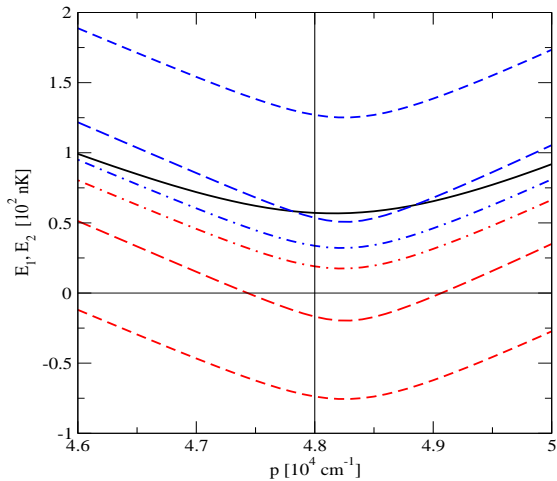,height=8cm,width=9cm,angle=0}
\caption{ (Color online)
The dependence of the quasiparticle  spectra of two hyperfine 
states $E_1$ and $E_2$ on the momentum for $\alpha = 0=
\delta\epsilon$ (solid line);  $\alpha = 0.05$ and $\delta\epsilon =
0$ (dashed lines);  $\alpha = 0.05$, $\delta\epsilon = 0.1$, 
$x = 0$ (dashed-dotted) and $x = 1$ (short-dashed lines).
The Fermi-momentum $k_F = 4.8$ $10^4$ cm$^{-1}$ is indicated 
by the vertical line. The remaining parameters are as in Fig. 1.
}
\label{MSfig:fig2}
\end{figure}                   
Figure~\ref{MSfig:fig2} compares the quasiparticle spectra $E_1$ and 
$E_2$ for combinations of $\alpha$ and $\delta\epsilon$. An important 
feature of the asymmetric ($\alpha\neq 0$, $\delta\epsilon =  0$) 
spectrum is its gapless nature, i.e. the existence of nodes for one
(or both) branches of the spectra (c.f. with the gapped BCS spectrum 
also shown in Fig.~\ref{MSfig:fig2}). The spectra of the DFS phase cover 
the range bounded by the curves with $x = 0$ and  $x = 1$.  We
conclude that the spectrum of deformed superfluid states is likewise 
gapless for a range of the angles defined by the variable $x$.
The macroscopic features of the atomic DFS phase, 
such as responses to the density perturbations or electromagnetic
probes, and their thermodynamic functions (heat capacity, etc) 
would differ from the ordinary BCS  phase due to the nodes and 
anisotropy of their spectrum, as is the case for gapless 
and/or anisotropic metallic superconductors.
\begin{figure}[t] % fig 3
\epsfig{figure=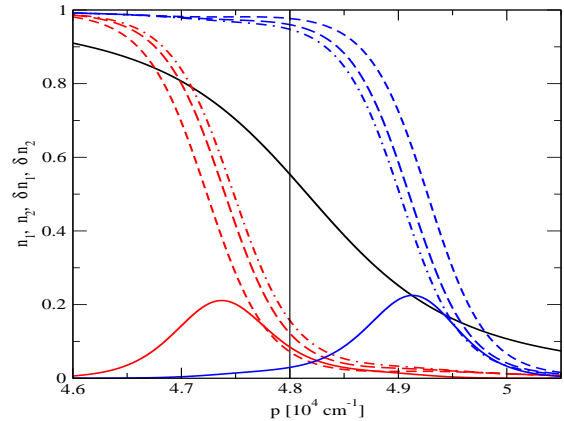,height=12cm,width=7cm,angle=-90}
\caption{   (Color online)
The dependence of the occupation probabilities of two hyperfine 
states on the momentum. The Fermi-momentum $k_F = 4.8$ $10^4$
cm$^{-1}$ is indicated by the vertical line. The labeling of 
the lines is as follows: $\alpha = 0=
\delta\epsilon$ (solid line),  $\alpha = 0.05$ and $\delta\epsilon =
0$ (dashed lines);  $\alpha = 0.05$, $\delta\epsilon = 0.1$, 
$x = 0$ (dashed-dotted) and $x = 1$ (short-dashed lines).
The bell-shaped curves show the anisotropy 
- the difference between the $x=1$ and $x=0$ occupation numbers -
for $\alpha = 0.05$, $\delta\epsilon = 0.1$.
The remaining parameters are as in Fig. 1.
}

\label{MSfig:fig3}
\end{figure} 
Figure~\ref{MSfig:fig3} shows the occupation numbers in BCS, 
deformed and/or asymmetric cases; varying the cosine of the 
polar angle $x$ covers a range of probabilities which includes 
the undeformed asymmetric state. The bell-shaped curves show 
the angular polarization of the occupation numbers defined as
$\delta n_{\sigma} = \vert n_{\sigma}(x=1) - n_{\sigma}(x=0)\vert$.
We observe up to $20\%$ anisotropy in the occupation 
probabilities of particles along and orthogonal to the symmetry 
breaking axis.

In closing, we would like to address the issue of an
experimental detection of the DFS phase. A direct way
to detect the DFS phase is the measurement of 
the anisotropy in the momentum distribution of the trapped atoms. 
Such a measurement can be realized by the time-of-flight technique
\cite{MOLECULAR_BEC,JILA04}.
This method uses the fact that after releasing the trap,
the atoms fly out freely and an image of their spatial distribution
taken after some time of flight  provides information on 
their momentum distribution when confined inside the trap. 
Assuming that the system was in the deformed superfluid state
one would  detect a mean momentum of
particles of type $1$ (majority) in the direction of symmetry 
breaking by about 20\% larger than that of particles of
type $2$ (minority) in the same direction.
Therefore, the presence of anisotropy in the detected momentum
distributions is an evidence for a deformed {\it superfluid} state
being the ground state of the system, as deformation alone
({\it i. e.} without pairing) would not lower the energy so as to
produce a deformed non-superfluid ground state.
The direction of spontaneous symmetry breaking (in $k$-space and,
therefore, also in real space) is chosen by the system randomly
and needs to be located in an experiment to obtain maximum anisotropy.

The Barcelona-T\"ubingen collaboration was supported in part by the 
programs DGICYT No. BFM2002-01868 and HA2001-0023 [Acciones
Integradas] (Spain) and the Deutsche  Akademische Austausch Dienst 
and Sonderforschungsberich 381 of the Deutsche Forschungsgemeinschaft 
(Germany). J. M.-P. would like to acknowledge the
participants of the Levico Workshop on Ultracold Fermi gases, and
particularly R. Combescot and N. Nygaard, for valuable discussions.
J. M.-P. also acknowledges the support from the Generalitat de Catalunya. 

%------------------------------------------------------------------------------

%\newpage
\end{document}